\documentclass[proceedings]{JHEP3}
\usepackage{eurosym}
\usepackage{amssymb}
\usepackage{amsfonts}
\usepackage{amsmath}
\usepackage{epsfig}

\newbox\mybox

\newcommand\fverb{\setbox\mybox=\hbox\bgroup\verb}
\newcommand\fverbdo{\egroup\medskip\noindent\fbox{\unhbox\mybox}\ }
\newcommand\fverbit{\egroup\item[\fbox{\unhbox\mybox}]}
\conference{Solvable 2D quantum systems with $\infty$D Hilbert space in the broken PT-regime}
\abstract{We provide exact analytical solutions for a two dimensional explicitly time-dependent non-Hermitian quantum system. While the time-independent 
variant of the model studied is in the broken PT-symmetric phase for the entire range of the model parameters, and has therefore a partially complex energy eigenspectrum,
its time-dependent version has real energy expectation values at all times. In our solution procedure we compare the two equivalent approaches of directly solving the time-dependent 
Dyson equation with one employing the Lewis-Riesenfeld method of invariants. We conclude that the latter approach simplifies the solution procedure due to the fact that the invariants of the non-Hermitian and Hermitian system
are related to each other in a pseudo-Hermitian fashion, which in turn does not hold for their corresponding time-dependent Hamiltonians. Thus constructing invariants and subsequently 
using the pseudo-Hermiticity relation between them allows to compute the Dyson map and to solve the Dyson equation indirectly. In this way one can bypass to solve
nonlinear differential equations, such as the dissipative Ermakov-Pinney equation emerging in our and many other systems.}

\title{Solvable two dimensional time-dependent non-Hermitian quantum systems
with infinite dimensional Hilbert space in the broken PT-regime}
\author{Andreas Fring and Thomas Frith \\
Department of Mathematics, City, University of London,\\
Northampton Square, London EC1V 0HB, UK\\
E-mail: a.fring@city.ac.uk, thomas.frith@city.ac.uk}

\input{tcilatex}
\begin{document}

\section{Introduction}

In the context of non-Hermitian time-independent quantum mechanics many
systems are known to posses real spectra in a certain parameter regime that
becomes spontaneously broken when some coupling constants are driven beyond
the exceptional point \cite{Bender:1998ke,Benderrev,Alirev,moiseyev2011non}.
Unlike their optical analogues \cite{Muss,MatMakris,Guo}, where the
spontaneously broken regime is of great interest, in quantum mechanics this
regime is usually discarded on grounds of being nonphysical since it leads
inevitably to infinite growth in energy due to the fact that the energy
eigenvalues emerge as complex conjugate pairs. In \cite{AndTom3} we
demonstrated that the introduction of an explicit time-dependence into a
non-Hermitian Hamiltonian can make the spontaneously broken $\mathcal{PT}$%
-regime physically meaningful. The reason for this phenomenon is that the
energy operator becomes modified due an additional term related to the Dyson
operator and hence its expectation values can become real. Here we extend
the previous analysis of the broken $\mathcal{PT}$-regime from a one
dimensional two-level system \cite{AndTom3} to a two-dimensional system with
infinite Hilbert space.

In addition, we show that technically it is simpler to employ
Lewis-Riesenfeld invariants \cite{Lewis69} instead of directly solving the
time-dependent Dyson map or the time-dependent quasi-Hermiticity relation.
All approaches are of course equivalent, but the invariant method splits the
problem into several more treatable steps. In particular, it can be viewed
as reformulating the nonpseudo-Hermitian relation for the Hamiltonians
involved, i.e. the time-dependent Dyson relation, into a pseudo-Hermitian
relation for the corresponding invariants. The latter quantities are well
studied in the time-independent setting and are far easier to solve as they
do not involve derivatives with respect to time. Loosely speaking the
time-derivative in the time-dependent Dyson relation acting on the Dyson map
has been split up into the two time-derivatives acting on the invariants
ensuring their conservation. Besides this aspect related to the
technicalities associated to the solution procedure we also provide the
first explicitly solved time-dependent system in higher dimensions.

Our manuscript is organized as follows: In section 2 we recall the key
equations that determine the Dyson map and hence the metric operator. In
section 3 we introduce our two-dimensional model. As first we demonstrate
how it may be solved in a time--independent setting. Subsequently we
determine the time-dependent Dyson map in two alternative ways, comparing
the direct and the Lewis-Riesenfeld method. In addition, we compute the
analytical solutions to the time-dependent Schr\"{o}dinger equation and use
them to evaluate instantaneous energy expectation values. Our conclusions
are stated in section 4.

\section{Time-dependent Dyson equation versus Lewis-Riesenfeld invariants}

The central object to compute in the study non-Hermitian Hamiltonian systems
is the metric operator $\rho $ that can be expressed in terms of the Dyson
operator $\eta $ as $\rho =\eta ^{\dagger }\eta $. Unlike as in the
time-independent scenario a non-Hermitian Hamiltonian $H(t)\neq H^{\dagger
}(t)$ can no longer be related to a Hermitian counterpart $h(t)=h^{\dagger
}(t)$ in a pseudo-Hermitian way, that is via a similarity transformation,
but instead the two Hamiltonians are related to each other by means of the
time-dependent Dyson relation 
\begin{equation}
h(t)=\eta (t)H(t)\eta ^{-1}(t)+i\hbar \partial _{t}\eta (t)\eta ^{-1}(t).
\label{hH}
\end{equation}%
When the Hamiltonian $h(t)$ is observable, this relation implies immediately
that the Hamiltonian $H(t)$ is not observable \cite%
{CA,time1,time6,fringmoussa} as it is not a self-adjoint operator with
regard to the standard or modified inner product. The Hamiltonians are
understood to be the operators governing the time-evolution of the systems
satisfying the time-dependent Schr\"{o}dinger equations 
\begin{equation}
\mathcal{H}(t)\Psi _{\mathcal{H}}(t)=i\hbar \partial _{t}\Psi _{\mathcal{H}%
}(t),\qquad \text{for }\mathcal{H}=h,H.  \label{TS}
\end{equation}%
The Hamiltonian is only identical to the observable energy operator in the
Hermitian case, but different in the non-Hermitian setting where it has to
be modified to 
\begin{equation}
\tilde{H}(t):=\eta ^{-1}(t)h(t)\eta (t)=H(t)+i\hbar \eta ^{-1}(t)\partial
_{t}\eta (t).  \label{Henergy}
\end{equation}%
The two wavefunctions in (\ref{TS}) are related to each other by the Dyson
map%
\begin{equation}
\Psi _{h}(t)=\eta (t)\Psi _{H}(t).  \label{sol}
\end{equation}%
Besides the time-dependent Dyson relation also the time-dependent
quasi-Hermiticity relation is then modified, by acquiring an additional
derivative term in the metric operator%
\begin{equation}
H^{\dagger }(t)\rho (t)-\rho (t)H(t)=i\hbar \partial _{t}\rho (t).
\label{qH}
\end{equation}

It was demonstrated \cite{fringmoussa,fringmoussa2,AndTom1,AndTom2,AndTom3}
that the equations (\ref{hH}) and (\ref{qH}) can be directly solved
consistently for $\eta (t)$ and $\rho (t)$, respectively. Alternatively, but
completely equivalent, one may also employ the standard Lewis-Riesenfeld
approach \cite{Lewis69} of computing invariants as argued in \cite%
{khantoul2017invariant,maamache2017pseudo}. This approach requires to
compute the two conserved time-dependent invariants $I_{h}(t)$ and $I_{H}(t)$%
, i.e. $dI_{h}/dt=dI_{H}/dt=0$, from the evolution equations 
\begin{equation}
\partial _{t}I_{\mathcal{H}}(t)=i\hbar \left[ I_{\mathcal{H}}(t),\mathcal{H}%
(t)\right] ,\qquad ~~~\ \ \text{for~\ }\mathcal{H}=h=h^{\dagger },H\neq
H^{\dagger }.  \label{LR0}
\end{equation}%
Using these two equations together with the Dyson relation (\ref{hH}) it is
straightforward to derive that the two invariants are simply related by a
similarity transformation 
\begin{equation}
I_{h}(t)=\eta (t)I_{H}(t)\eta ^{-1}(t)\text{.}  \label{simhH}
\end{equation}%
Since the invariant $I_{h}$ is Hermitian, the invariant $I_{H}$ is its
pseudo-Hermitian counterpart. When $I_{h}$ and $I_{H}$ have been
constructed, (\ref{simhH}) is a much easier equation to solve for $\eta (t)$%
, than directly the Dyson relation (\ref{hH}). At this point one has
therefore also obtained the metric operator simply by $\rho =\eta ^{\dagger
}\eta $. Next one may also employ the invariants to construct the
time-dependent eigenstates from the standard equations \cite{Lewis69} 
\begin{eqnarray}
~~~I_{\mathcal{H}}(t)\left\vert \phi _{\mathcal{H}}(t)\right\rangle 
&=&\Lambda \left\vert \phi _{\mathcal{H}}(t)\right\rangle ,~~~~~~~~~~~~~~\ \
\ \ \ \ \ ~\ ~\left\vert \Psi _{\mathcal{H}}(t)\right\rangle =e^{i\hbar
\alpha (t)}\left\vert \phi _{\mathcal{H}}(t)\right\rangle ,~~~~~  \label{LR1}
\\
\dot{\alpha} &=&\left\langle \phi _{\mathcal{H}}(t)\right\vert i\hbar
\partial _{t}-\mathcal{H}(t)\left\vert \phi _{\mathcal{H}}(t)\right\rangle
,\qquad \dot{\Lambda}=0~  \label{LR2}
\end{eqnarray}%
for $\mathcal{H}=h$ and $\mathcal{H}=H$. Below we compare the two approaches
and conclude that even though the approach using invariants is more lengthy,
it dissects the original problem into several easier smaller steps when
compared to solving the Dyson equation directly. Of course both approaches
are equivalent and must lead to the same solutions for $\eta (t)$, as we
also demonstrate.

In what follows we set $\hbar =1$.

\section{2D systems with infinite Hilbert space in the broken $\mathcal{PT}$%
-regime}

\subsection{Two dimensional time-independent models}

We set up our model by considering at first a $\mathcal{PT}$-symmetric
system that we then slightly modify by going from a model with partially
broken $\mathcal{PT}$-symmetry to one with completely broken $\mathcal{PT}$%
-symmetry. We commence with one of the simplest options for a
two-dimensional non-Hermitian system by coupling two harmonic oscillators
with a non-Hermitian coupling term in space 
\begin{equation}
H_{xy}=\frac{1}{2m}\left( p_{x}^{2}+p_{y}^{2}\right) +\frac{1}{2}m\left(
\Omega _{x}^{2}x^{2}+\Omega _{y}^{2}y^{2}\right) +i\kappa xy,~~~~~~m,\kappa
,\Omega _{x},\Omega _{y}\in \mathbb{R}.  \label{Napkin}
\end{equation}%
This non-Hermitian Hamiltonian is symmetric with regard to the antilinear
transformations \cite{EW} $\mathcal{PT}_{\pm }:x\rightarrow \pm x$, $%
y\rightarrow \mp y$, $p_{x}\rightarrow \mp p_{x}$, $p_{y}\rightarrow \pm
p_{y}$, $i\rightarrow -i$, i.e. $\left[ \mathcal{PT}_{\pm },H_{xy}\right] =0$%
. Using standard techniques from $\mathcal{PT}$-symmetric/quasi-Hermitian
quantum mechanics \cite{Bender:1998ke,Benderrev,Alirev}, it can be decoupled
easily into two harmonic oscillators 
\begin{equation}
h_{xy}=\eta H_{xy}\eta ^{-1}=\frac{1}{2m}\left( p_{x}^{2}+p_{y}^{2}\right) +%
\frac{1}{2}m\left( \omega _{x}^{2}x^{2}+\omega _{y}^{2}y^{2}\right) ,
\end{equation}%
by a simple rotation using the angular momentum operator $L_{z}=xp_{y}-yp_{x}
$ in the Dyson map $\eta =e^{\theta L_{z}}$ and constraining the parameters
involved as%
\begin{equation}
\omega _{x}^{2}=\frac{\Omega _{x}^{2}\cosh ^{2}\theta +\Omega _{y}^{2}\sinh
^{2}\theta }{\cosh 2\theta },~~\omega _{y}^{2}=\frac{\Omega _{x}^{2}\sinh
^{2}\theta +\Omega _{y}^{2}\cosh ^{2}\theta }{\cosh 2\theta },~~\tanh
2\theta =\frac{2\kappa }{m\left( \Omega _{y}^{2}-\Omega _{x}^{2}\right) }.
\label{xx}
\end{equation}%
By the last equation in (\ref{xx}) it follows that one has to restrict $%
\left\vert \kappa \right\vert \leq m\left( \Omega _{y}^{2}-\Omega
_{x}^{2}\right) /2$ for this transformation to be meaningful. Thus as long
as the Dyson map is well defined, i.e. the constraint holds, the energy
eigenspectra 
\begin{equation}
E_{n,m}=\left( n+\frac{1}{2}\right) \omega _{x}+\left( m+\frac{1}{2}\right)
\omega _{y}.
\end{equation}%
of $h$ and $H$ are identical and real. The restriction on $\kappa $ is the
same as the one found in \cite{MandalMY,beygi2015}, where the decoupling of $%
H$ to $h$ was realized by an explicit coordinate transformation instead of
the Dyson map. In fact, identifying the parameter $k$ in \cite{MandalMY} as $%
k=\cosh 2\theta $, and somewhat similarly in \cite{beygi2015}, the
coordinate transformation becomes a rotation realized by the similarity
transformation acting on the coordinates and the momenta, i.e. we obtain $%
H\rightarrow h$ with the coordinate transformation%
\begin{equation}
v\rightarrow ~~\eta v\eta ^{-1}=\left( 
\begin{array}{cc}
\cosh \theta  & i\sinh \theta  \\ 
-i\sinh \theta  & \cosh \theta 
\end{array}%
\right) v,~~~~~\text{for }v=\left( 
\begin{array}{c}
x \\ 
y%
\end{array}%
\right) ,\left( 
\begin{array}{c}
p_{x} \\ 
p_{y}%
\end{array}%
\right) .
\end{equation}%
Such a scenario is mostly well understood and in analogy to the case studied
in \cite{AndTom3}, solving the time-dependent Dyson equation for $\eta (t)$
will allow to make sense of the regime for $\kappa \rightarrow \kappa (t)$
beyond the exceptional point.

Let us now slightly modify the model above by modifying some of the
constants and by adding a term that also couples the two harmonic oscillator
Hamiltonians in the momenta%
\begin{equation}
H_{xyp}=\frac{a}{2}\left( p_{x}^{2}+x^{2}\right) +\frac{b}{2}\left(
p_{y}^{2}+y^{2}\right) +i\frac{\lambda }{2}\left( xy+p_{x}p_{y}\right)
,\qquad a,b,\lambda \in \mathbb{R}.  \label{xyp}
\end{equation}%
Clearly this Hamiltonian is also symmetric with regard to the same
antilinear symmetry as $H_{xy}$, i.e. we have $\left[ \mathcal{PT}_{\pm
},H_{xyp}\right] =0$. Thus we expect the eigenvalues to be real or to be
grouped in pairs of complex conjugates when the symmetry is broken for the
wavefunctions.

It is convenient to express this Hamiltonian in a more generic algebraic
fashion as 
\begin{equation}
H_{K}=aK_{1}+bK_{2}+i\lambda K_{3},  \label{Hk}
\end{equation}%
where we defined Lie algebraic generators 
\begin{equation}
K_{1}=\frac{1}{2}\left( p_{x}^{2}+x^{2}\right) ,~~K_{2}=\frac{1}{2}\left(
p_{y}^{2}+y^{2}\right) ,~~K_{3}=\frac{1}{2}\left( xy+p_{x}p_{y}\right)
,~~K_{4}=\frac{1}{2}\left( xp_{y}-yp_{x}\right) .  \label{om}
\end{equation}%
Besides the generators already appearing in the Hamiltonian we added one
more generator, $K_{4}=L_{z}/2$, to ensure the closure of the algebra, i.e.
we have%
\begin{equation}
\begin{array}{lll}
\left[ K_{1},K_{2}\right] =0,~ & \left[ K_{1},K_{3}\right] =iK_{4}, & \left[
K_{1},K_{4}\right] =-iK_{3}, \\ 
\left[ K_{2},K_{3}\right] =-iK_{4},~~ & \left[ K_{2},K_{4}\right] =iK_{3},~~
& \left[ K_{3},K_{4}\right] =i(K_{1}-K_{2})/2.%
\end{array}
\label{alg}
\end{equation}%
Notice that $K_{i}^{\dagger }=K_{i}$ for $i=1,\ldots ,4$. In what follows we
mostly use the algebraic formulation so that our results also hold for
representations different from (\ref{om}). We report that the Hamiltonian $%
H_{xy}$ in (\ref{Napkin}) requires at least a ten dimensional Lie algebra
when demanding $xy$ to be one of the Lie algebraic generators, which is the
reason we consider first the more compactly expressible Hamiltonian $H_{xyp}$%
.

Using the same form of the Dyson map $\eta =e^{\theta L_{z}}$ as above,
albeit with $\theta =\func{arctanh}[\lambda /(b-a)]$, this Hamiltonian is
decoupled into 
\begin{equation}
h_{K}=\eta H_{K}\eta ^{-1}=\frac{1}{2}(a+b)\left( K_{1}+K_{2}\right) +\frac{1%
}{2}\sqrt{(a-b)^{2}-\lambda ^{2}}\left( K_{1}-K_{2}\right) ,
\end{equation}%
for $\left\vert \lambda \right\vert <\left\vert a-b\right\vert $. So clearly
for $a=b$ we are in the completely broken $\mathcal{PT}$-regime. That choice
is in addition very convenient as it allows for a systematic construction of
the eigenvalue spectrum of $H_{K}(b=a)$. Since the following commutators
vanish $\left[ H_{K}(b=a),K_{1}+K_{2}\right] =$ $\left[ H_{K}(b=a),K_{3}%
\right] =\left[ K_{1}+K_{2},K_{3}\right] =0$, one simply needs to search for
simultaneous eigenstates of $K_{3}$ and $K_{1}+K_{2}$ to determine the
eigenstates if $H_{K}(b=a)$, due to Schur's lemma. Indeed for the
representation (\ref{om}) we obtain for $H_{K}(b=a)$ the eigenstates%
\begin{equation}
\varphi _{n,m}(x,y)=\frac{e^{-\frac{x^{2}}{2}-\frac{y^{2}}{2}}}{2^{n+m}\sqrt{%
n!m!\pi }}\left[ \dsum\limits_{k=0}^{n}\binom{n}{k}H_{k}(x)H_{n-k}(y)\right] %
\left[ \dsum\limits_{l=0}^{m}(-1)^{l}\binom{m}{l}H_{l}(y)H_{m-l}(x)\right] ,
\end{equation}%
with corresponding eigenenergies%
\begin{equation}
E_{n,m}=E_{m,n}^{\ast }=a(1+n+m)+i\frac{\lambda }{2}(n-m).
\end{equation}%
Here $H_{n}(x)$ denotes the $n$-th Hermite polynom in $x$. The states are
orthonormal with regard to the standard inner product $\left\langle \varphi
_{n,m}\right. \left\vert \varphi _{n^{\prime },m^{\prime }}\right\rangle
=\delta _{n,n^{\prime }}\delta _{m,m^{\prime }}$. The reality of the
subspectrum with $n=m$ is explained by the fact that the $\mathcal{PT}_{\pm }
$-symmetry is preserved, i.e. we can verify that $\mathcal{PT}_{\pm }$ $%
\varphi _{n,n}=\varphi _{n,n}$. However, when $n\neq m$ the $\mathcal{PT}%
_{\pm }$-symmetry is spontaneously broken and the eigenvalues occur in
complex conjugate pairs.

Hence this Hamiltonian should be discarded as nonphysical in the
time-independent regime, but we shall see that it becomes physically
acceptable when the parameters $a$ and $\lambda $ are taken to be explicitly
time-dependent.

\subsection{A solvable 2D time-dependent Hamiltonian in the broken $\mathcal{%
PT}$-regime}

We solve now the explicitly time-dependent non-Hermitian Hamiltonian 
\begin{equation}
H(t)=\frac{a(t)}{2}\left( p_{x}^{2}+p_{y}^{2}+x^{2}+y^{2}\right) +i\frac{%
\lambda (t)}{2}\left( xy+p_{x}p_{y}\right) ,\qquad a(t),\lambda (t)\in 
\mathbb{R}.  \label{H}
\end{equation}%
According to the above discussion, the instantaneous eigenvalue spectrum of $%
H(t)$ belongs to the spontaneously broken $\mathcal{PT}$-regime.

\subsubsection{The time-dependent Dyson equation}

Let us now compute the right hand side of the time-dependent Dyson relation (%
\ref{hH}). For that purpose we assume that the Dyson map is an element of
the group associated to the algebra (\ref{alg}) and take it to be of the form%
\begin{equation}
\eta (t)=\dprod\nolimits_{i=1}^{4}e^{\gamma _{i}(t)K_{i}},\qquad \gamma
_{i}\in \mathbb{R}.  \label{eta}
\end{equation}%
As $\eta $ is not a unitary operator by definition, we have taken the $%
\gamma _{i}$ to be real to avoid irrelevant phases. Using now (\ref{eta})
and (\ref{H}) in (\ref{hH}), the right hand side will be Hermitian if and
only if%
\begin{equation}
\gamma _{1}=\gamma _{2}=q_{1},\quad \dot{\gamma}_{3}=-\lambda \cosh \gamma
_{4},\quad \dot{\gamma}_{4}=\lambda \tanh \gamma _{3}\sinh \gamma _{4},
\label{34}
\end{equation}%
for some real constant $q_{1}\in \mathbb{R}$. The Hermitian Hamiltonian
results to 
\begin{equation}
h(t)=a(t)\left( K_{1}+K_{2}\right) +\frac{\lambda (t)}{2}\frac{\sinh \gamma
_{4}}{\cosh \gamma _{3}}\left( K_{1}-K_{2}\right) .  \label{hher}
\end{equation}%
For the representation (\ref{om}) these are simply two decoupled harmonic
oscillators with time-dependent coefficients. The energy operator $\tilde{H}$
as defined in equation (\ref{Henergy}) becomes%
\begin{equation}
\tilde{H}(t)=a(t)\left( K_{1}+K_{2}\right) +\frac{\lambda (t)}{4}\sinh
(2\gamma _{4})\left( K_{1}-K_{2}\right) -i\lambda (t)\left( \sinh ^{2}\gamma
_{4}K_{3}-\sinh \gamma _{4}\tanh \gamma _{3}K_{4}\right) .
\end{equation}

The constraining relations (\ref{34}) may be solved directly for $\gamma
_{3} $ and $\gamma _{4}$, but not in a straightforward manner.\ We eliminate 
$\lambda $ and $dt$ from the last two equations in (\ref{34}), so that $%
d\gamma _{4}=-\tanh \gamma _{3}\tanh \gamma _{4}d\gamma _{3}$, hence
obtaining $\gamma _{4}$ as a function of $\gamma _{3}$ 
\begin{equation}
\gamma _{4}=\func{arcsinh}\left( \kappa \func{sech}\gamma _{3}\right)
\label{43}
\end{equation}%
with integration constant $\kappa $. Defining $\chi (t):=\cosh \gamma _{3}$
we use (\ref{34}) and (\ref{43}) to derive that the central equation that
needs to be satisfied is the Ermakov-Pinney equation \cite{Ermakov,Pinney}
with a dissipative term%
\begin{equation}
\ddot{\chi}-\frac{\dot{\lambda}}{\lambda }\dot{\chi}-\lambda ^{2}\chi =\frac{%
\kappa ^{2}\lambda ^{2}}{\chi ^{3}}.  \label{DEP}
\end{equation}%
This equation is ubiquitous in the context of solving time-dependent
Hermitian systems, even in the Hermitian setting, see e.g. \cite%
{leach2008ermakov}. While some solutions to this equation are known, we
demonstrate here that solving this nonlinear differential equation can be
completely bypassed when employing Lewis-Riesenfeld invariants instead and
computing $\eta $ from the pseudo-Hermiticity relation (\ref{simhH}) for the
invariants instead.

\subsubsection{The time-dependent Dyson map from pseudo-Hermiticity}

It is natural to assume that the invariants $I_{H}$, $I_{h}$ as well as the
Hermitian Hamiltonian $h(t)$ lie in the same algebra as the non-Hermitian
Hamiltonian $H(t)$. Furthermore we note that $I_{h}(t)$ needs to be
Hermitian, so that we make the Ans\"{a}tze%
\begin{equation}
I_{H}(t)=\dsum\limits_{i=1}^{4}\alpha _{i}(t)K_{i},~~\ \
~~~I_{h}(t)=\dsum\limits_{i=1}^{4}\beta _{i}(t)K_{i},\quad
~~h(t)=\dsum\limits_{i=1}^{4}b_{i}(t)K_{i},  \label{IhH}
\end{equation}%
with~$\alpha _{i}=\alpha _{i}^{r}+i\alpha _{i}^{i}\in \mathbb{C}$, $%
b_{i},\beta _{i},\alpha _{i}^{r},\alpha _{i}^{i}\in \mathbb{R}$.

\paragraph{The Lewis-Riesenfeld invariant $I_{H}(t)$:}

Substituting the expressions for $I_{H}(t)$ and $H(t)$ into the equation in (%
\ref{LR0}) and reading off the coefficients of the generators $K_{i}$ we
obtain the four constraints%
\begin{equation}
\dot{\alpha}_{1}=\frac{i}{2}\lambda \alpha _{4},\quad ~\dot{\alpha}_{2}=-%
\frac{i}{2}\lambda \alpha _{4},\quad ~\dot{\alpha}_{3}=0,\quad ~\dot{\alpha}%
_{4}=i\lambda (\alpha _{2}-\alpha _{1}).
\end{equation}%
These equations are easily solved by%
\begin{equation}
\alpha _{1}=\frac{c_{1}}{2}+c_{3}\cosh \left[ c_{4}-\!\!\dint\limits_{0}^{t}%
\lambda (s)ds\right] ,~~\alpha _{2}=c_{1}-\alpha _{1},~~\alpha
_{3}=c_{2},~~\alpha _{4}=2ic_{3}\sinh \left[ c_{4}-\!\!\dint\limits_{0}^{t}%
\lambda (s)ds\right] ,  \label{alpha}
\end{equation}%
with complex integration constants $c_{i}=c_{i}^{r}+ic_{i}^{i}$, $%
c_{i}^{r},c_{i}^{i}\in \mathbb{R}$. At this point we have two options, we
may either compute directly the invariant $I_{h}(t)$ for the Hamiltonian $%
h(t)$ as given in (\ref{hher}) by using the evolution equation (\ref{LR0})
or the similarity relation (\ref{simhH}) instead.

\paragraph{The Lewis-Riesenfeld invariant $I_{h}(t)$:}

Denoting the coefficients of $K_{1}$ and $K_{2}$ in (\ref{hher}) by $b_{1}(t)
$ and $b_{2}(t)$, respectively, as defined in the expansion for generic $h(t)
$ in (\ref{IhH}), the relation for the invariants (\ref{LR0}) leads to the
constraints%
\begin{equation}
\dot{\beta}_{1}=0,\quad ~\dot{\beta}_{2}=0,\quad ~\dot{\beta}_{3}=\beta
_{4}(b_{2}-b_{1}),\quad ~\dot{\beta}_{4}=\beta _{3}(b_{1}-b_{2}).
\end{equation}%
These four coupled first order differential equations are easily solved by%
\begin{equation}
\beta _{1}=c_{5},\quad \beta _{2}=c_{6},\quad \beta _{3}=c_{7}\cos \left[
c_{8}-\!\!\dint\nolimits_{0}^{t}(b_{1}-b_{2})ds\right] ,\quad \beta
_{4}=-c_{7}\sin \left[ c_{8}-\!\!\dint\nolimits_{0}^{t}(b_{1}-b_{2})ds\right]
.  \label{sol1}
\end{equation}%
Next we invoke the pseudo-Hermiticity relation for the invariants (\ref%
{simhH}).

\paragraph{Relating $I_{H}(t)$ and $I_{h}(t)$:}

So far we have treated the Hermitian and non-Hermitian systems separately.
Next we relate them using the Ans\"{a}tze (\ref{eta}) for $\eta (t)$ and (%
\ref{IhH}) for the invariants in the expression (\ref{simhH}). We obtain
eight equations by reading off the coefficients and separating the result
into real and imaginary parts. We can solve the resulting equations for the
real functions 
\begin{eqnarray}
\beta _{1} &=&\frac{1}{2}\left[ \alpha _{1}^{r}+\alpha _{2}^{r}-\alpha
_{4}^{i}\sinh \gamma _{3}+\alpha _{3}^{i}\sinh \gamma _{4}\cosh \gamma
_{3}+\left( \alpha _{1}^{r}-\alpha _{2}^{r}\right) \cosh \gamma _{3}\cosh
\gamma _{4}\right] ,  \label{sol2} \\
\beta _{2} &=&\frac{1}{2}\left[ \alpha _{1}^{r}+\alpha _{2}^{r}+\alpha
_{4}^{i}\sinh \gamma _{3}-\alpha _{3}^{i}\sinh \gamma _{4}\cosh \gamma
_{3}-\left( \alpha _{1}^{r}-\alpha _{2}^{r}\right) \cosh \gamma _{3}\cosh
\gamma _{4}\right] , \\
\beta _{3} &=&\left( \alpha _{2}^{i}-\alpha _{1}^{i}\right) \sinh \gamma
_{4}+\alpha _{3}^{r}\cosh \gamma _{4}, \\
\beta _{4} &=&\left[ \left( \alpha _{1}^{i}-\alpha _{2}^{i}\right) \cosh
\gamma _{4}-\alpha _{3}^{r}\sinh \gamma _{4}\right] \sinh \gamma _{3}+\alpha
_{4}^{r}\cosh \gamma _{3}
\end{eqnarray}%
with the additional constraints%
\begin{eqnarray}
\alpha _{1}^{i}+\alpha _{2}^{i} &=&0,~~~~~~\ \ \ \ \ \ \ \ \ \ \ \ \ \alpha
_{3}^{r}\alpha _{3}^{i}+\alpha _{4}^{r}\alpha _{4}^{i}=2\alpha
_{1}^{i}(\alpha _{2}^{r}-\alpha _{1}^{r}),~~  \label{con} \\
\tanh \gamma _{3} &=&\frac{\alpha _{4}^{i}}{\sqrt{\left( \alpha
_{1}^{r}-\alpha _{2}^{r}\right) ^{2}-(\alpha _{3}^{i})^{2}}},~~~~~~\tanh
\gamma _{4}=\frac{\alpha _{3}^{i}}{\alpha _{2}^{r}-\alpha _{1}^{r}}~~.
\label{con2}
\end{eqnarray}%
We also used here $\gamma _{1}=\gamma _{2}$.

Next we compare our solutions in (\ref{alpha}), (\ref{sol1}) and (\ref{sol2}%
)-(\ref{con2}). First we use the expressions for the $\alpha _{i}$ from (\ref%
{alpha}) in (\ref{sol2})-(\ref{con2}). The constraints (\ref{con}) imply
that $c_{1}^{i}=0~$and $4c_{3}^{r}c_{3}^{i}=-c_{2}^{r}c_{2}^{i}$ so that the
time-dependent coefficients in the Hermitian invariant $I_{h}$ result to%
\begin{eqnarray}
\beta _{1} &=&\frac{c_{1}^{r}}{2}\pm \frac{1}{2}\sqrt{%
4(c_{3}^{r})^{2}-(c_{2}^{i})^{2}},  \label{v1} \\
\beta _{2} &=&\frac{c_{1}^{r}}{2}\pm \frac{1}{2}\sqrt{%
4(c_{3}^{r})^{2}-(c_{2}^{i})^{2}},  \label{v2} \\
\beta _{3} &=&\pm \frac{c_{2}^{r}}{2c_{3}^{r}}\frac{\left[
4(c_{3}^{r})^{2}-(c_{2}^{i})^{2}\right] }{\sqrt{%
4(c_{3}^{r})^{2}-(c_{2}^{i})^{2}\func{sech}^{2}\left[ c_{4}^{r}-\dint%
\nolimits_{0}^{t}\lambda (s)ds\right] }},  \label{v3} \\
\beta _{4} &=&\pm \frac{c_{2}^{r}c_{2}^{i}}{2c_{3}^{r}}\sqrt{\frac{%
4(c_{3}^{r})^{2}-(c_{2}^{i})^{2}}{4(c_{3}^{r})^{2}-(c_{2}^{i})^{2}\func{sech}%
^{2}\left[ c_{4}^{r}-\dint\nolimits_{0}^{t}\lambda (s)ds\right] }}\tanh %
\left[ c_{4}^{r}-\dint\nolimits_{0}^{t}\lambda (s)ds\right] ,  \label{v4}
\end{eqnarray}%
with the constraint $2\left\vert c_{3}^{r}\right\vert >\left\vert
c_{2}^{i}\right\vert $. These expressions need to match with those computed
directly in (\ref{sol2}). It is clear how to identify the constants $c_{5}$
and $c_{6}$ in (\ref{sol1}) when comparing to (\ref{v1}) and (\ref{v2}).
Less obvious is the comparison between the $\beta _{3}$ and $\beta _{4}$.
Reading off $b_{1}$ and $b_{2}$ from (\ref{hher}) and using (\ref{con2}), we
compute%
\begin{equation}
\dint\nolimits_{0}^{t}(b_{1}-b_{2})ds=\arctan \left[ \frac{c_{2}^{i}}{\sqrt{%
4(c_{3}^{r})^{2}-(c_{2}^{i})^{2}}}\tanh \left[ c_{4}^{r}-\dint%
\nolimits_{0}^{t}\lambda (s)ds\right] \right] .
\end{equation}%
Setting next the constants $c_{8}=0$, $c_{7}=\pm c_{2}^{r}\sqrt{%
4(c_{3}^{r})^{2}-(c_{2}^{i})^{2}}/(2c_{3}^{r})$ the solution in (\ref{sol1})
matches indeed with (\ref{v3}) and (\ref{v4}).

We can now assemble our expressions for $\eta $ by using the results for $%
\gamma _{3}$ and $\gamma _{4}$ from (\ref{con2}) together with the
expressions in (\ref{alpha}) obtaining%
\begin{eqnarray}
\gamma _{3} &=&\arctan \left[ \frac{\tanh \left[ q_{2}-\dint%
\nolimits_{0}^{t}\lambda (s)ds\right] }{\sqrt{1-q_{3}^{2}\func{sech}\left[
q_{2}-\dint\nolimits_{0}^{t}\lambda (s)ds\right] ^{2}}}\right] ~,~~
\label{g34} \\
\gamma _{4} &=&-\func{arccot}\left[ \frac{1}{q_{3}}\cosh \left[
q_{2}-\dint\nolimits_{0}^{t}\lambda (s)ds\right] \right] ,
\end{eqnarray}%
with the identification $q_{2}=c_{4}^{r}$ and $q_{3}=c_{2}^{i}/(2c_{3}^{r})$.

We convince ourselves that the function 
\begin{equation}
\chi (t)=\cosh \gamma _{3}=\sqrt{\frac{\cosh ^{2}\left[ q_{2}-\dint%
\nolimits_{0}^{t}\lambda (s)ds\right] -q_{3}^{2}}{1-q_{3}^{2}}}
\label{solEP}
\end{equation}%
computed with $\gamma _{3}$ as given in (\ref{g34}) does indeed satisfy the
dissipative Ermakov-Pinney equation (\ref{DEP}) when identifying the
constants as $\kappa =q_{3}/\sqrt{1-q_{3}^{2}}$. We also express the
Hamiltonian (\ref{hher}) explicitly as%
\begin{equation}
h(t)=f_{+}(t)K_{1}+f_{-}(t)K_{2}~~~~\text{with }f_{\pm }(t)=a(t)\pm \frac{%
q_{3}\sqrt{1-q_{3}^{2}}\lambda (t)}{1+\cosh \left[ 2q_{2}-2\dint%
\nolimits_{0}^{t}\lambda (s)ds\right] -2q_{3}^{2}},  \label{fpm}
\end{equation}%
which is evidently Hermitian for $\left\vert q_{3}\right\vert <1$.

\subsubsection{Eigenstates, phases and instantaneous energy expectation
values}

We note that the computation of the Dyson map did not require the knowledge
of any eigenstates, neither when using Lewis-Riesenfeld invariants nor in
the directly approach of solving the time-dependent Dyson relation. This
also means that so far we have not solved the time-dependent Schr\"{o}dinger
equation nor did we use the eigenstate equations (\ref{LR1}) and (\ref{LR2}%
). Let us therefore carry out the final step and determine all eigenstates,
including relevant phases, and use them to evaluate the energy expectation
values.

The exact solution to the time-dependent Schr\"{o}dinger equation for the
harmonic oscillator with time-dependent mass and frequency is well known for
twenty years \cite{pedrosa1997exact}. Adapting that solution here to our
notation and situation, for the Hamiltonian $\tilde{h}(t)=$ $a(t)K_{1}$,
with $a(t)$ being any real function of $t$, it reads%
\begin{equation}
\tilde{\varphi}_{n}(x,t)=\frac{e^{i\alpha _{n}(t)}}{\sqrt{\varkappa (t)}}%
\exp \left[ \left( \frac{i}{a(t)}\frac{\dot{\varkappa}(t)}{\varkappa (t)}-%
\frac{1}{\varkappa ^{2}(t)}\right) \frac{x^{2}}{2}\right] H_{n}\left[ \frac{x%
}{\varkappa (t)}\right] ,~~\   \label{Ped}
\end{equation}%
with phase 
\begin{equation}
\alpha _{n}(t)=-\left( n+\frac{1}{2}\right) \ \dint\nolimits_{0}^{t}\frac{%
a(s)}{\varkappa ^{2}(s)}ds,
\end{equation}%
and $\varkappa (t)$ being restricted to the dissipative Ermakov-Pinney
equation 
\begin{equation}
\ddot{\varkappa}-\frac{\dot{a}}{a}\dot{\varkappa}+a^{2}\varkappa =\frac{a^{2}%
}{\varkappa ^{3}}.  \label{EP2}
\end{equation}%
Thus while we could bypass to solve this equation in the form of (\ref{DEP})
for the determination of $\eta $ when it involved $\lambda $, it has
re-emerged for the computation of the eigenstates involving $a$ with a
different sign in front of the last term on the right hand side. Using the
wavefunction (\ref{Ped}) we compute here the expectation value for $K_{1}$
and a normalization factor 
\begin{eqnarray}
\left\langle \tilde{\varphi}_{n}(x,t)\right\vert K_{1}\left\vert \tilde{%
\varphi}_{m}(x,t)\right\rangle  &=&2^{n-2}n!(2n+1)\sqrt{\pi }\frac{%
a^{2}(1+\varkappa ^{4})+\varkappa ^{2}\dot{\varkappa}^{2}}{a^{2}\varkappa
^{2}}\delta _{n,m},  \label{ex1} \\
\left\langle \tilde{\varphi}_{n}(x,t)\right. \left\vert \tilde{\varphi}%
_{n}(x,t)\right\rangle  &=&2^{n}n!\sqrt{\pi }:=N.
\end{eqnarray}%
Next we notice that the expectation value (\ref{ex1}) does not depend on
time 
\begin{equation}
\frac{d}{dt}\left[ \frac{a^{2}(1+\varkappa ^{4})+\varkappa ^{2}\dot{\varkappa%
}^{2}}{a^{2}\varkappa ^{2}}\right] =\frac{2\dot{\varkappa}}{a^{2}}\left( 
\ddot{\varkappa}-\frac{\dot{a}}{a}\dot{\varkappa}+a^{2}\varkappa -\frac{a^{2}%
}{\varkappa ^{3}}\right) =0.  \label{EPf}
\end{equation}%
by recognizing in (\ref{EPf}) one of the factors as the Ermakov-Pinney
equation in the form (\ref{EP2}). It is clear that this constant will
dependent on the explicit solution for (\ref{EP2}). So for definiteness we
compute it by adapting the solution (\ref{solEP}) to account for the
aforementioned different sign 
\begin{equation}
\varkappa (t)=\sqrt{\tilde{\kappa}\cos \left[ 2\dint\nolimits_{0}^{t}a(s)ds%
\right] +\sqrt{1+\tilde{\kappa}^{2}}}.  \label{solk}
\end{equation}%
For this solution we calculate%
\begin{equation}
\frac{a^{2}(1+\varkappa ^{4})+\varkappa ^{2}\dot{\varkappa}^{2}}{%
a^{2}\varkappa ^{2}}=2\sqrt{1+\tilde{\kappa}^{2}}.
\end{equation}%
Thus for the normalized wavefunction $\hat{\varphi}_{n}(x,t)=\tilde{\varphi}%
_{m}(x,t)/\sqrt{N}$ involving the solution (\ref{solk}) we find%
\begin{equation}
\left\langle \hat{\varphi}_{n}(x,t)\right\vert K_{1}\left\vert \hat{\varphi}%
_{m}(x,t)\right\rangle =\left( n+\frac{1}{2}\right) \sqrt{1+\tilde{\kappa}%
^{2}}\delta _{n,m}.
\end{equation}%
Hence the solution to the time-dependent Schr\"{o}dinger equation for the
Hermitian Hamiltonian $h(t)$ in (\ref{fpm}) is simply 
\begin{equation}
\Psi _{h}^{n,m}(x,y,t)=\hat{\varphi}_{n}^{+}(x,t)\hat{\varphi}_{m}^{-}(y,t)
\end{equation}%
when replacing $a\rightarrow f^{\pm }$, $\varkappa \rightarrow \varkappa
_{\pm }$, $\tilde{\kappa}\rightarrow \tilde{\kappa}_{\pm }$ and $\alpha
_{n}\rightarrow \alpha _{n}^{\pm }$ in an obvious manner. We have now
assembled all the information needed to compute the instantaneous energy
expectation values%
\begin{eqnarray}
E^{n,m}(t) &=&\left\langle \Psi _{h}^{n,m}(t)\right\vert h(t)\left\vert \Psi
_{h}^{n,m}(t)\right\rangle =\left\langle \Psi _{H}^{n,m}(t)\right\vert \rho
(t)\tilde{H}(t)\left\vert \Psi _{H}^{n,m}(t)\right\rangle   \label{Eexp} \\
&=&f_{+}(t)\left( n+\frac{1}{2}\right) \sqrt{1+\tilde{\kappa}_{+}^{2}}%
+f_{-}(t)\left( m+\frac{1}{2}\right) \sqrt{1+\tilde{\kappa}_{-}^{2}},  \notag
\end{eqnarray}%
with constants $\kappa _{\pm }$. It is clear that this expectation value is
real for any given time-dependent fields $a(t)$, $\lambda (t)\in \mathbb{R}$
and constants $\tilde{\kappa}_{\pm }\in \mathbb{R}$, $\left\vert
q_{3}\right\vert <1$. Hence, we have explicitly shown that one can draw the
same conclusion as in the one-dimensional case \cite{AndTom3}, that a
time-independent non-Hermitian Hamiltonian in the spontaneous spontaneously
broken $\mathcal{PT}$-regime becomes physically meaningful in the
time-dependent\ setting.

\section{Conclusions}

We have presented the first higher dimensional solution of the
time-dependent Dyson relation (\ref{hH}) relating a non-Hermitian and a
Hermitian Hamiltonian system with infinite dimensional Hilbert space. As for
the one dimensional case studied in \cite{AndTom3}, we have demonstrated
that the time-independent non-Hermitian system in the spontaneously broken $%
\mathcal{PT}$-regime becomes physically meaningful when including an
explicit time-dependence into the parameters of the model and allowing the
metric operator also to be time-dependent. The energy operator (\ref{Henergy}%
) has perfectly well-defined real expectation values (\ref{Eexp}).

Technically we have compared two equivalent solution procedures, solving the
time-dependent Dyson relation directly for the Dyson map or alternatively
computing Lewis-Riesenfeld invariants first and subsequently constructing
the Dyson map from the similarity relation that related the Hermitian and
non-Hermitian invariants. The latter approach was found to be simpler as the
similarity relation is far easier than the differential version (\ref{hH}).
The price one pays in this approach is that one needs to compute the two
invariants first. However, the differential equations for these quantities
turned out to be easier than the (\ref{hH}). In particular, it was possible
to entirely bypass the dissipative Ermakov-Pinney equation in the
computation of $\eta (t)$. Nonetheless, this ubiquitous equation re-emerged
in the evaluation of the eigenfunctions involving different time-dependent
fields and with a changed sign.

\bigskip \noindent \medskip \textbf{Acknowledgments:} TF is supported by a
City, University of London Research Fellowship.

\end{document}